\documentstyle[12pt,epsf]{article}
\begin{document}

\title{Numerical Study on Space-Time\\ Pulse Compression}
\author{Monika Pietrzyk}
\date {\small \sl Institute of Fundamental Technological Research \\
Polish Academy of Sciences \\
 \'Swi\c etokrzyska 21, 00-049 Warsaw, Poland}
\maketitle

\begin{abstract}

A numerical study of the 
properties of Gaussian pulses propagating in planar waveguide 
under the combined effect of positive Kerr-type 
nonlinearity, diffraction  in planar waveguides 
and anomalous or normal dispersion, is presented. 
It is demonstrated how the relative strength of dispersion 
and diffraction, the strength of nonlinearity and the initial spatial 
and temporal pulse chirps effect on the parameters of pulse compression, 
 such as the maximal compression factor and the distance to the point 
of maximal compression.
\end{abstract}
\pagebreak
	
\section{Introduction}
\normalsize
A compression of optical pulses in Kerr-type nonlinear media have
 been subject to investigation for many years 
and continues to attract a certain attention \cite{compr4, compr5}. 
In single-mode fibers with anomalous group-velocity 
dispersion (GVD) and positive nonlinearity the pulse compression is based 
on the mechanism of higher-order soliton generation \cite {spm3}.
 In single-mode fibers with normal GVD the pulse compression can be 
obtained in the configuration with a grating pair \cite {spm2, spm6}. 
In both cases the self-phase modulation (SPM) induced by an intense pulse is  
used. However, the intense pump pulse propagating together with a weak probe 
pulse can also cause pulse compression by the 
mechanisms of the so-called cross-phase modulation 
(XPM) \cite {XPM}, or the  induced-phase modulation (IPM) \cite {ipm}.

 A possibility of pulse compression in non-dispersive nonlinear bulk media
due to another nonlinear effect, that of self-focusing,  
is discussed in \cite {coll2}-\cite{coll4} with the aid  of the paraxial ray 
approximation, \cite{coll2, coll4}, 
and by means of the variational analysis, \cite{coll3}. 
Still another pulse compression technique that uses 
the self confinement of two-dimensional spatial bright solitons 
propagating in non-dispersive bulk media is 
mentioned in \cite {twosol},  where the two-beam interference technique is 
used in order to ensure that a filamentation 
(a splitting of the beam into many sub-beams) does not  occur. 

Moreover, a simultaneous space-time collapse, which can occur in  bulk
media and  in  planar waveguides under the combined effect of 
nonlinearity, diffraction and anomalous dispersion, may also be useful 
for pulse compression \cite {sf1,coll5}. This kind of collapse gives rise to 
short pulses with extremely high optical field  \cite {Var1, coll1, super}.
It   is realizable both in the
case when dispersion and diffraction have comparable effect on pulse 
propagation and  in the more general case when 
one of the effects above is dominating (see \cite {compr3}).

On the other side,  the interplay of  normal dispersion and positive 
nonlinearity causes quite different behavior of the pulse.  
In optical fibers where diffraction terms are not included 
it leads to a monotonic pulse spreading. 
However, the inclusion of the diffraction term, which is 
necessary  for a planar waveguide, can lead to a pulse
compression, as it was described  in \cite{compr3, compr2}.
Besides, in planar waveguide, normal dispersion slows the 
self-focusing of the pulse and causes a splitting of the pulse 
into two pulses \cite{compr2, Darek}. The effect of splitting of a pulse was
observed also in the bulk media \cite{sf1}.

In this paper a compression of a pulse propagating in  planar, self-focusing
 nonlinear planar waveguide in the regime of  anomalous and   normal 
dispersion is considered. The structure of the paper is as follows. 
In Section 2, the nonlinear Schr\"odinger equation 
describing dispersive pulse propagation in nonlinear planar waveguides 
and the parameters of pulse compression are introduced. 
In Section 3, an estimation of the condition of pulse collapse 
is made with the aid of the so-called method of moments \cite{moment2}. 
Numerical results describing the influence of the magnitude of  
nonlinearity, the relative strength of dispersion and diffraction 
and the spatial and temporal chirp of the initial Gaussian pulse 
on the pulse compression parameters are discussed in Section 4.
 
\section{Basic equations}
It is well known that starting from the Maxwell equations for the 
envelope $U(x,y,z,t)$ of the electric field 
\[ E(x,y,z,t)=U(x,y,z,t)e^{-i(\omega t-n_0\beta _0 z)} \] 
 propagating along the $z$ axis in a planar waveguide  
with positive, instantaneous Kerr-type nonlinearity, 
one obtains the 2-dimensional nonlinear Schr\"odinger equation (NSE) 
 \cite{compr2} :
\normalsize
\begin{equation}
  i \frac {\partial} {\partial \zeta} U -\frac 1 2 \sigma \frac{\partial ^2}
{\partial \tau  ^2} U + \frac 1 2 \frac {\partial ^2} { \partial \xi ^2} U 
+ N^2 \mid U \mid^2 U =0 , 
\label{nse}
\end{equation}
\normalsize
if the paraxial and the 
slowly varying envelope approximations are applied 
and the term $\nabla  \cdot (\nabla E)$, 
the shock term \cite {splitting} proportional to 
$\frac{\partial (|E|^2E)}{\partial t}$ 
and higher-order dispersion effects can be neglected. 

In Equation 1,	
$ \zeta = \frac z {z_f}$ is the normalized longitudinal spatial coordinate,
$\xi = \frac x {w_0}$ is the normalized transverse spatial coordinate,
$\tau = \frac{t-\beta _1 z}{t_0}$ is the normalized local time,
$\sigma = \frac {\beta _2 z_f}{t_0^2}$ represents the relative strength of
 dispersion and diffraction, 
$N=\beta_0 U_0 w_0 \sqrt{n_0 n_2}$ parameterizes  the strength of nonlinearity,
$\beta_0=\frac {\omega} c $ is a wave number,
$\beta_n=\frac{d^n \beta}{d\omega^n}$ are dispersion terms,
$z_f = \beta_0 n_0 w_0^2 $ is the Fresnel diffraction length,
$w_0$ is the spatial width of the input pulse,
$t_0$ is the temporal width of the input pulse (i.e., duration of the 
input pulse), 
$U_0 $ is the peak amplitude of the input pulse, and  
$n=n_0+n_2|U|^2$ is the refraction index for the Kerr type nonlinear media.
Recall that $\sigma>0 $ corresponds to  the normal dispersion  
and $\sigma<0$ corresponds to the anomalous dispersion.

 As the initial condition we take the  Gaussian chirped pulse
which is given by (cf. \cite{moment6}) 
\begin{equation}
U(\xi,\tau,\zeta =0)=
e^{-\frac{\xi ^2 (1+ i C_{\xi})}{2}}
e^{-\frac{\tau ^2 (1 + i C_{\tau} )}{2}},
\end{equation}
where $C_{\xi}$ $(C_{\tau})$ is the spatial (temporal) pulse chirp
(the focusing spatial chirp corresponds to $C_{\xi}<0$ and  
the focusing temporal chirp corresponds to $sgn(- \sigma C_{\tau} )\\ <0$).

We will characterize a pulse by its spatial width, $w_{\xi}(\zeta)$, and the 
temporal width, $w_{\tau}(\zeta)$, which are defined by 
\[ 
U(w_{\xi},0,\zeta)=\frac 1 e U(0,0,\zeta) \;
\mbox{\hspace{0.3cm} \rm  and\hspace*{0.3cm}} \; 
U(0,w_{\tau},\zeta)=\frac 1 e U(0,0,\zeta).
\] 
We also introduce the maximal compression factor 
\[
c_{max}=\frac {\tau_0} {w_{\tau \, min}(\zeta_{m})} , 
\] 
 where $w_{\tau \, min}(\zeta_m)$ 
is the minimal temporal width of the pulse (see \cite {compr2, compr1}).
In following we call $\zeta_m$  the position of the minimal pulse width.
 
Solution of NSE (Equation 1) with the initial condition given by Equation 2
can describe a propagation of a dispersive Gaussian pulse 
in nonlinear planar waveguides. It is worth remarking that for the anomalous 
dispersion regime a solutions of this equation can also describe a 
dispersion-less elliptic Gaussian beam, \cite{astigmatic, elliptical}
(i.e. a cw beam with elliptic Gaussian 
transverse profile) propagating in a nonlinear bulk media. 

In this paper we refer to the case of $\sigma = -1$ as  the 
cylindrically symmetric spatiotemporal pulse; 
the case of $\sigma \neq -1$ is to be referred to as the 
asymmetric spatiotemporal pulse.

In the particular case of  the cylindrical spatiotemporal pulse
a simple analytic solution of the NSE exists 
which describes a behavior of beam propagating in nonlinear media
by means of the variational approximation \cite{Var1} or 
by means of the scaled complex rays formulation within  
the so-called ABCD matrix formalism (see \cite{ray, abcd}). 
For the asymmetric spatiotemporal pulse only a semi-analytical approach  
of \cite {astigmatic, elliptical} is known in the literature.

It is known that some  solutions of the two- or three-dimensional NSE 
can develop into a singularity of the electric field when the initial 
pulse power exceeds a certain critical value \cite{Var1}. 
This phenomenon, known as  
a pulse collapse, 
 can occur simultaneously in space and time for a pulse propagating in planar 
waveguide with the anomalous GVD \cite {Var1, coll1}, 
and also  for a 
dispersion-less beam propagating in self-focusing bulk medium. 
This singularity, however, is obviously non-physical, 
for it emerges just as an artifact  
of the paraxial approximation made when deriving the NSE.
In order to avoid this  
limitation, either a non-paraxial treatment of the 
process of self-focusing 
\cite {non} or some other effects, such as the nonlinear absorption 
and the saturation of the nonlinear refractive index, 
should be taken into consideration.
From another hand, the appearance of a non-physical singularity in 
numerical simulations based on NSE can serve 
as an indication to the real collapse taking place in the certain point 
of space. This is in fact the criterion used in Section 4.

Studying the details of developing the pulse collapse we leave  
beyond the scope of this paper.
Instead, our task is to determine the  values of the parameters $\sigma$ and 
$N^2$ for which the pulse collapse can occur. 
For this purpose the so-called method of moments \cite {moment4} could be 
used. However,  it gives 
only an estimation of the sufficient conditions of the pulse collapse,
whereas the latter can occur, in fact, 
 at the earlier times or on the shorter propagation distances \cite {coll4}. 
More precise conditions will be obtained by means of the 
numerical simulations presented in Section 3, (cf. also \cite{Var1}) 

\section{Sufficient conditions of pulse collapse}

The method of moments originates from the paper of Vlasov e.a. \cite {moment4}.
It can be used as an approach to the determination 
of whether a given initial wave pulse can collapse to a singular point 
in a finite period of time \cite {moment1}. 
An application of the method of moments
to the NSE may be found in \cite {moment2}.

In order to formulate the condition of collapse 
in terms of the strength of nonlinearity, $N^2$, and  
the relative strength of dispersion and diffraction, $\sigma$, 
we first introduce the second moment of intensity
\[
I(\zeta)=\int\limits_{-\infty}^{\infty} \int\limits_{-\infty}^{\infty}
(\xi ^2 +\bar{\tau }^2) |U| ^2 d\xi d\bar {\tau},
\]
where $U$ is a solution of the NSE given by Equation 1, with the normalization
$\bar {\tau}=(-\sigma) ^{-\frac 1 2} \tau$,  ($\sigma \neq 0$). 

Parameter $I$ can be interpreted as effective beam size measuring
the size of the area to which most of the energy is confined. 

Assuming that $U$ decay suitably as $r \to \infty $, 
one can obtain \cite{moment2}
\begin{equation}
\frac{d^2 I}{d \zeta^2}= \ddot I =4E
\end{equation}
where $E$ is the Hamiltonian of the NSE
\[
E=\int \int \left (\frac 1 2 \left |\frac {\partial U(\xi, \bar {\tau})}
{\partial \xi} \right |^2 
+\frac 1 2 \left | \frac {\partial U(\xi, \bar {\tau})}
{\partial \bar {\tau}}\right | ^2 
-\frac 1 2 N^2 |U(\xi, \bar {\tau})|^4 \right )d\xi d\bar {\tau}
\]
Because $E$ remains constant during a pulse propagation, i.e. it is 
independent of $\zeta$, Equation 3 may be integrated twice to give :
\[
I(\zeta)=2E\zeta ^2 + \dot I(0)\zeta +I(0),
\]
where $\dot I = \frac {dI} {d\zeta}$. 

If the right-hand side of the above equation vanished, then the pulse width 
(both spatial and temporal) will decrease to zero in a finite distance 
leading to beam collapse. Therefore a sufficient condition for collapse
can occur if the following conditions are satisfied 
\cite {moment2, moment6, moment1}:
\begin {eqnarray}
&E<0& ,\nonumber\\
&E=0& \hspace{0.5cm} \mbox{and} \hspace {0.5cm} \dot I(0)<0 ,\\
&E>0& \hspace{0.5cm} \mbox{and} \hspace{0.5cm} \dot I(0) < -\sqrt{8EI(0)}.
\nonumber
\end{eqnarray}

For Gaussian input pulse, given by Equation 2, Hamiltonian $E$
can be expressed in the following form 
\[ 
E= \frac 1 2 \int \int \left [ 
\left ( \xi^2 - \sigma^2 \bar {\tau}^2 \right ) e^{-\xi^2}
e^{-\sigma \bar{\tau}^2}-N^2 
e^{-2\xi^2}e^{-2\sigma \bar{\tau}^2}\right ] d\xi d\bar{\tau}=
\frac 1 4 \sqrt{-\sigma} \pi (1-\sigma -N^2).
\]

In the particular case of flat phase front, $C_{\xi}=C_{\tau}=0$, 
we obtain that $\dot I(0)=0$ and because of this two last 
criterion in Equation 4 are not satisfied. The first criterion, 
$E<0$, yields 
\begin{equation}
N^2>1-\sigma.
\end {equation}

Equation 5 may be considered as the sufficient condition 
of the pulse collapse in terms of  the strength of nonlinearity,
$N^2$, which is proportional to the peak amplitude, $|U_0|^2$,
and the relative strength of dispersion and diffraction, $\sigma$. 
The magnitude of the parameter $N^2$ which is sufficient for the pulse
 collapse to occur increases linearly with $|\sigma|$. 
This is not unexpected because the collapse 
of the pulse occurs when the self-focusing caused by the nonlinearity 
dominates over the broadening of a pulse,
which is due to the diffraction and dispersion.   
It is obvious that for smaller values of the parameter $|\sigma|$ is,
the influence of the dispersion on the pulse broadening is weaker.

Note that the sufficient conditions of pulse collapse can  be formulated 
also in terms of the critical initial power $P_c$ of the pulse as follows
\cite{moment6, elliptical} 
\[ 
\frac{P_c}{P_0}=\left [\sqrt{|\sigma|}+\frac 1{\sqrt{|\sigma|}}\right ] 
\geq 1 ,
\]
where $P_c(\sigma)=\int |U|^2d \bar {\tau} d \xi =\pi |U_0|^2 \frac {1}
 {\sqrt{|\sigma|}}$, $P_0=2 \pi$ is the initial power 
of the cylindrically symmetric pulse (i.e. $\sigma = 1$).

We conclude that the decrease of the parameter $|\sigma|$ 
leads to the decrease of the critical amplitude, $|U_0|^2$ , and to the 
increase of the critical power, $P_c$.

Note, that the collapse criteria obtained with  the aid of the 
method of moments for the particular case of the spatiotemporal 
symmetric pulse agrees with the 
result of the variational approximation in \cite{coll4, Var1}. 

\section{Numerical results and discussion}

\normalsize

In this Section, the results of numerical solution of the 2+1 dimensional 
NSE by means of the well-known Split-Step Spectral Method (SSSM) \cite{split} 
with the two dimensional (2d) Fast Fourier Transform \cite{num} are presented.
The calculations on the two-dimensional grid with $512 \times 512$ points 
(transverse steps, $\Delta \xi = \Delta \tau = 0.08$) and with 
the longitudinal step depending on the nonlinearity so that 
for $N^2=1$, $\Delta \zeta = 0.01$, were performed.
 Because of the lack of spatial-temporal cylindrical symmetry of the problem 
 it is not possible to 
simplify calculations by reducing the 2d Fast Fourier Transform 
to the one dimensional Hankel Transform developed in \cite{hankel1, hankel2}. 
Several checks of our numerical procedure were made, which 
include a simulation of  beam 
propagation in the absence of  group-velocity dispersion ($\sigma=0$), 
repeated testing with different transverse grid and longitudinal step 
length,  and the monitoring of pulse energy during each simulation. 
The latter was kept constant with an error  $ < 0.00005$ 

As initial conditions  in numerical calculations we take Gaussian 
pulse given by Equation 2. 
First, for the case of anomalous dispersion regime
we will compare conditions of pulse collapse predicted by
method of moments 
with those obtained from numerical calculations. 
Further, with the aid of numerical calculations we will study 
influence of the strength of nonlinearity, relative strength of 
dispersion and diffraction and spatial and temporal pulse
chirps on pulse compression parameters. Above  analysis will be
perform both for anomalous as well as normal dispersion regime.

\subsection {The anomalous dispersion regime}

In this section the influence  of the parameters $\sigma$ 
and $N^2$ on the pulse collapse and compression will be considered.

In Fig. 1, a comparison of the conditions of the collapse of
pulse predicted by the method of moments
with those obtained by numerical calculations is presented.
 In our numerical procedure the occurrence of  pulse collapse 
was identified  with the discontinuity of the phase $\phi(0,0,\zeta))$ 
in the central point of the pulse $u=|u|e^{i\phi}$, 
and with a non-monotonic behavior of the 
intensity in the central point of the pulse after reaching the 
collapse point.
The  results of numerical simulation are plotted by two kinds of points 
corresponding to the cases when, respectively,
the  pulse collapse  occurs or does not occur.
The prediction of the method of moments is given by the  
straight line $N^2=1-\sigma$, (see Equation 5). 
The boundary line between the collapse and the 
no collapse regions, obtained from the numerical data is 
approximately described by $N^2 \approx 0.85 - \sigma $.
It  is parallel to the straight line predicted by the 
method of moments, unless the absolute value of $\sigma$ is too small.

Therefore, for both methods, the magnitude of the parameter $N^2$
which is sufficient for the pulse collapse to occur increase 
linearly with $|\sigma|$. The discrepancy  
appears due to the theoretical idealization of the picture of the collapse 
where all the energy of the pulse goes to the singularity point. 
This also explains why conditions of numerical collapse are 
typically softer than those predicted by the method of moments described
in Section 3. 

Studying details of pulse collapse we leave beyond task of this
paper. Instead, we will study the influence of the relative 
strength of dispersion 
and diffraction, the nonlinearity and the spatial and temporal chirps
on the parameters of pulse compression
under the condition that the pulse collapse does not occur. 

Fig. 2 and Fig. 3 represent the results of calculations of the influence 
of the relative strength of dispersion and diffraction, $\sigma$, 
on the maximal compression factor, $c_{max}$, and on the position 
of the minimal pulse width, $\zeta_m$, for different values 
of the strength of nonlinearity,
$N^2$ and for Gaussian initial pulse with flat phase front. 
As it could be expected the parameters of pulse compression, 
$c_{max}$ and $\zeta_m$,increase monotonically 
with the increase of  $N^2$ and the decrease
of  $\sigma$ until collapse conditions are reached. 
 This behavior is obvious from the fact that 
increase of $N^2$ cause increase of pulse self-focusing, it helps
to concentrate pulse energy in the center, in addition, decrease of $\sigma$
cause decrease of dispersion broadening of the pulse. 

In Fig. 4 the results of numerical simulations   
of the influence of the initial spatial, $C_{\xi}=C$,
and two cases of temporal, $C_{\tau}= \pm C$,
chirps on the pulse compression parameters 
are presented.
In order to distinguish between the above two cases we introduce a parameter
\[ S=sgn (-C_{\xi} C_{\tau} \sigma),\]
which equals 1 for the case of focusing (defocusing) temporal
and spatial chirps and equals -1 for the case of focusing (defocusing)
temporal and defocusing (focusing) spatial chirps. 

As it could be expected, the focusing spatial and temporal chirps, 
$C<0, S=1$, cause the increase of the pulse compression parameters.
The explanation is that a defocusing chirp spreads the energy out 
from the center of the pulse, whereas a focusing chirp  
concentrates it there. As the result,  
the nonlinearity-induced phase curvature of the field is, 
respectively,  reduced  or enhanced. 
Similar effect of the focusing chirp
of the initial pulse takes place in the region close to the 
collapse. Namely,  the focusing spatial chirp can hasten the 
collapse, whereas a defocusing chirp can either delay or eliminate it 
entirely \cite{compr3}. 

More interesting is the case of $S=-1$
(i.e. the spatial focusing chirp and the temporal defocusing chirp occur 
simultaneously). The increase of the maximal compression factor
occurs only for the case of focusing temporal chirp,
$C>0$, whereas this is not 
always true for a spatial focusing chirp $C<0$, see Fig. 4. 
One can conclude that the temporal chirp has larger 
effect on the temporal pulse compression than he spatial one. One can expect
the reverse situation in the case of spatial compression of the pulse. 

\subsection {Normal dispersion regime}

In the case of normal dispersion regime 
the collapse of the pulse doesn't occur. 
However, due to the the spatiotemporal coupling occurring in nonlinear
medium when both the diffraction and the dispersion effects take place 
a pulse compression can be obtained \cite{compr2}. 

In this section 
we will study the influence of the relative strength of dispersion 
and diffraction, the nonlinearity and the spatial and temporal pulse chirps
on the parameters of pulse compression. 

It is seen from Fig. 5 that the maximal compression factor, $c_{max}$, 
monotonically decreases with $\sigma$, and increases with $N^2$.
It is clear because smaller value of the parameter 
$\sigma$ has a weaker influence on the dispersion broadening of the pulse,
moreover the increase of  $N^2$ leads to the increase of the spatiotemporal
coupling and nonlinearity induced  phase curvature of the field. 
In the end both effects lead to the temporal compression of the pulse.

From Fig. 6 it is seen that for sufficiently small values of $\sigma$ 
the parameter $\zeta_m$ decreases with $N^2$.
However it appears to be practically independent on $\sigma$ once 
a certain threshold level of $N^2$ is reached. 
This fact was  explained in \cite {compr3} by means of the periodic 
beam narrowing of  higher-order spatial solitons. 
A different behavior takes place at  larger values  of 
sigma ($\sigma > 0.25$). 
Namely, at first $\zeta_m$ increases with $N^2$ for sufficiently small 
$N^2$ and then it slowly decreases after reaching a maximal value at the 
certain value of $N^2$. 
This behavior is explainable by the fact that at small nonlinearities the 
effects of dispersion prevent a creation of spatial solitons.

In Fig. 7 the results of numerical calculations of the influence of the 
initial pulse chirp on the  parameters of pulse compression 
$c_m$ and $\zeta_m$
are presented. The focusing spatial and temporal chirps, $C<0, S=1$,
cause the increase of the compression parameters 
($c_{max}$ ans $\zeta_{max}$) 
and this behavior appears to be similar to that which we have 
 previously observed in Fig. 4 for the anomalous dispersion regime. 
However, in the case of the anomalous GVD   
$c_{max}$ grows with $C$ much faster that in the case of normal
GVD. Namely, for the anomalous GVD 
the maximal compression factor for a chirped initial pulse 
with $C=-2$ is three times larger than that for an  
initial pulse with flat phase front ($C=0$, i.e.  
$c_{max} (C=-2) = 3 \times c_{max}(C=0)$. 
For the normal GVD the increase of the $c_{max}$ is rather slow, e.g.  
$c_{max}(C=-2)= 1.1 \times c_{max}(C=0)$, 
and a saturation of the maximal compression 
factor occurs for the initial chirps below $-2$ (see Fig. 7).

Moreover, for the case of $C<0, S=-1$ 
(i.e. the spatial focusing chirp and the temporal defocusing chirp)
the maximal compression factor increases only for focusing 
temporal chirp, whereas this is not always true for a spatial 
focusing chirp $C_{\xi}<0$. 

\section{Conclusions}
\normalsize

In this paper, the physical conditions of collapse and compression 
of dispersive Gaussian pulses propagating in waveguide with 
the positive Kerr-type nonlinearity, diffraction and  the 
anomalous or normal dispersion are investigated. 

We determine the  values of the relative strength of dispersion and 
diffraction, $\sigma$,  and the strength of nonlinearity, $N^2$, 
for which the pulse collapse can occur. 
For this purpose we first present an estimation given by 
the method of moments \cite {moment4}. 
 More precise conditions are obtained by means of the numerical 
simulations based on the (2+1)-dimensional 
Nonlinear Schr\"odinger Equation (see Section 4).

We characterize a pulse compression by two parameters: 
the maximal compression factor, $c_{max}$,
and the distance to the point 
of the maximal compression, $\zeta_m$, (see Section 2).
By means of a numerical simulation 
we study how these two parameters depend on the parameters $N^2$ and 
$\sigma$, and the initial spatial and temporal pulse chirps. 
We demonstrate that in the regime of both  anomalous and normal 
dispersion  the increase of the nonlinearity and the decrease 
of the relative strength of dispersion and diffraction 
cause the increase of the maximal  compression factor. 
Moreover, in the case of anomalous dispersion regime the compression factor 
is maximal in the region of $1-\sigma \stackrel{<}{\sim}N^2$. 

Furthermore, we observe that the  increase of the focusing temporal and
spatial chirps of the initial pulse 
lead to the increase of the maximal compression factor, $c_{max}$. 
In the case of the anomalous GVD  
$c_{max}$ grows with chirp, $C$, much faster that 
in the case of normal GVD, for  which 
a saturation of the maximal compression factor occurs. 

Moreover, the increase of the focusing temporal chirp 
might lead, even in the presence of the defocusing spatial chirp,
 to the increase of the maximal compression factor, 
$c_{max}$, whereas  the defocusing temporal
chirp always leads to the  decrease of  $c_{max}$, 
even in the case of the focusing spatial chirp.  
It may be  concluded, therefore, that the temporal chirp has larger effect 
on the maximal pulse compression factor than the spatial chirp.
In reverse, it is expected that the spatial focusing chirp has larger
impact on the beam with than the temporal chirp, independently on
its sign. 

\section{Acknowledgments}
I would like to thank I. Kanatchikov for his helpful suggestions.

\pagebreak

\pagebreak

\begin{figure}
\raisebox{-1cm}[7cm][2cm]{
\hspace*{-2cm}
\epsffile{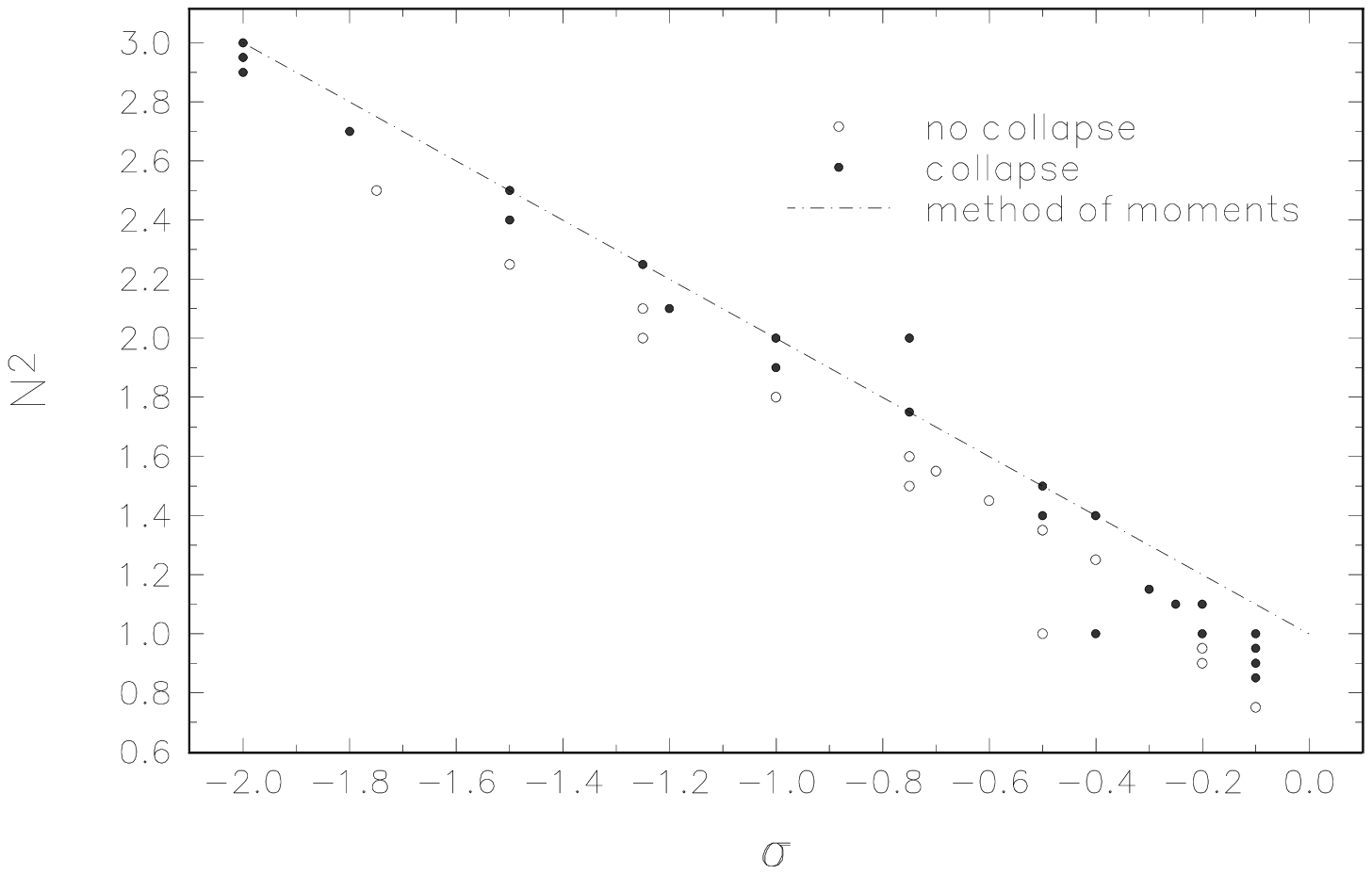}}
\caption{ Comparison of the sufficient conditions for pulse collapse 
predicted by the method of moments (straight line) and numerical 
calculations (filled circle points denote pulse collapse and 
empty circle points indicate no collapse). It was done for 
initial Gaussian pulse with flat phase front,
propagating in a medium described by two parameters: 
the strength of the nonlinearity, $N^2$ and 
the relative strength of dispersion and diffraction, $\sigma$.}
\label{gran}
\end{figure}

\begin{figure}
\raisebox{-1cm}[7cm][2cm]{
\hspace*{-2cm}
\epsffile{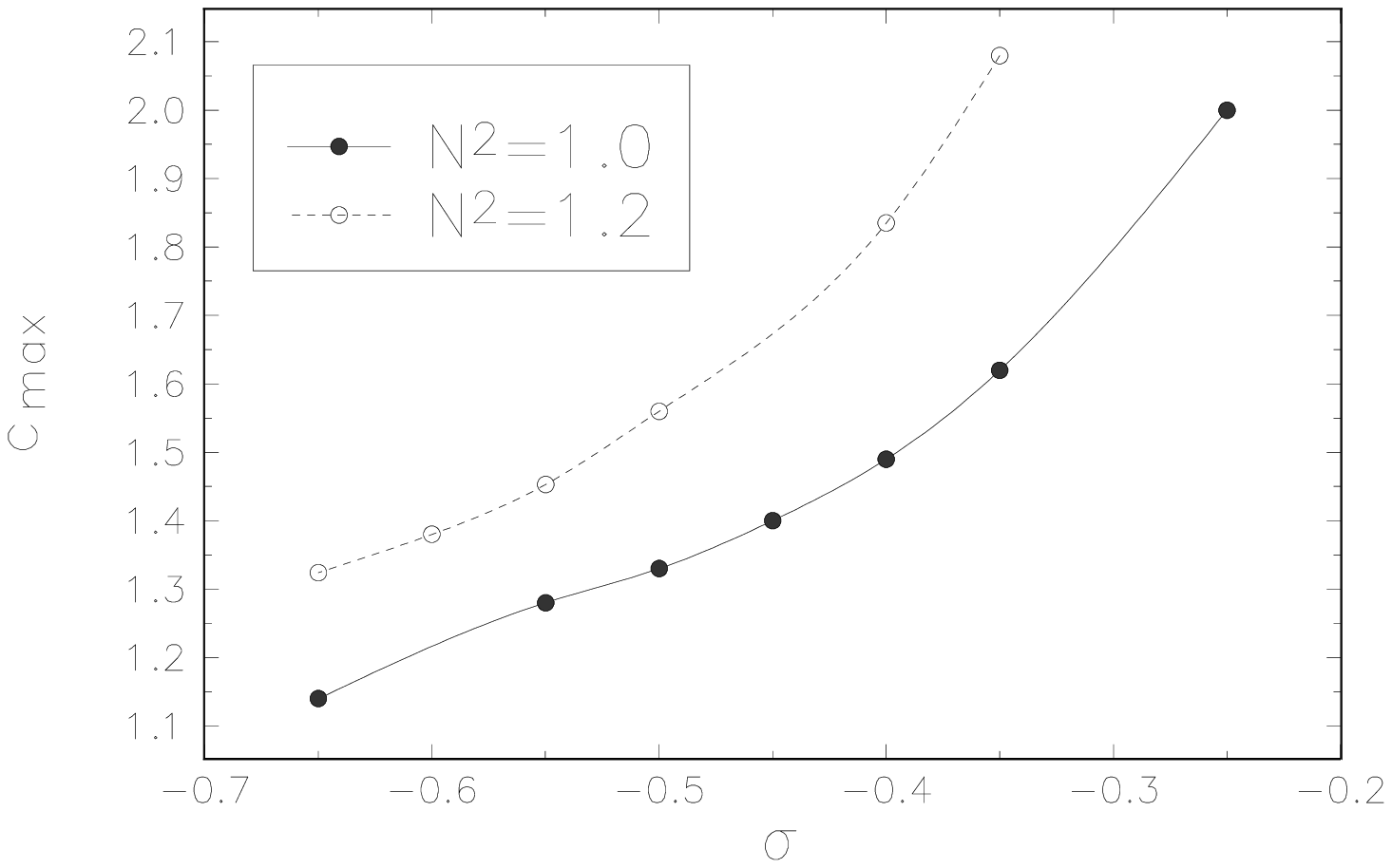}}
\caption{The maximum compression ratio, $c_{max}$ as a function 
of the relative strength of dispersion and  diffraction, $\sigma<0$, 
for different value of the strength of nonlinearity, $N^2$, 
and for initial Gaussian pulse with flat phase front.}
\label{scomp}
\end{figure}

\begin{figure}
\raisebox{-1cm}[7cm][2cm]{
\hspace*{-2cm}
\epsffile{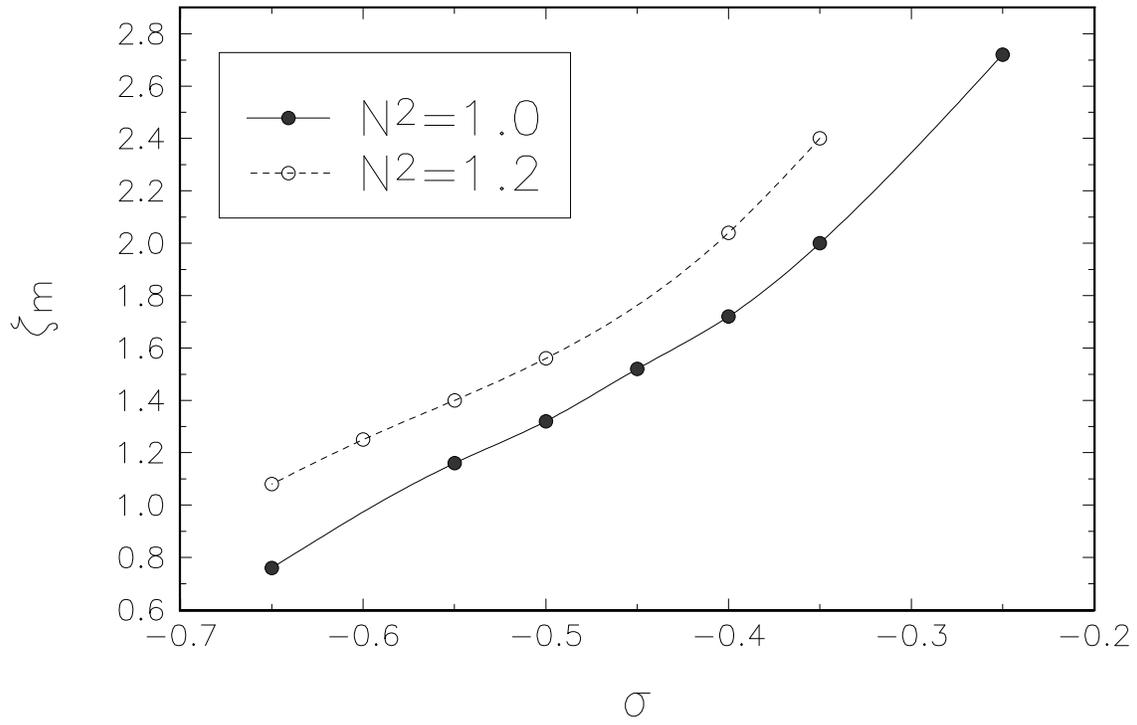}}
\caption{The distance to the point of minimal pulse width, 
$\zeta_m$, as a function of the relative strength of dispersion 
and diffraction, $\sigma<0$, for different value of the
strength of nonlinearity, $N^2$, and for initial Gaussian pulse 
with flat phase front.}
\label{sodl}
\end{figure}

\begin{figure}
\raisebox{-1cm}[7cm][2cm]{
\hspace*{-2cm}
\epsffile{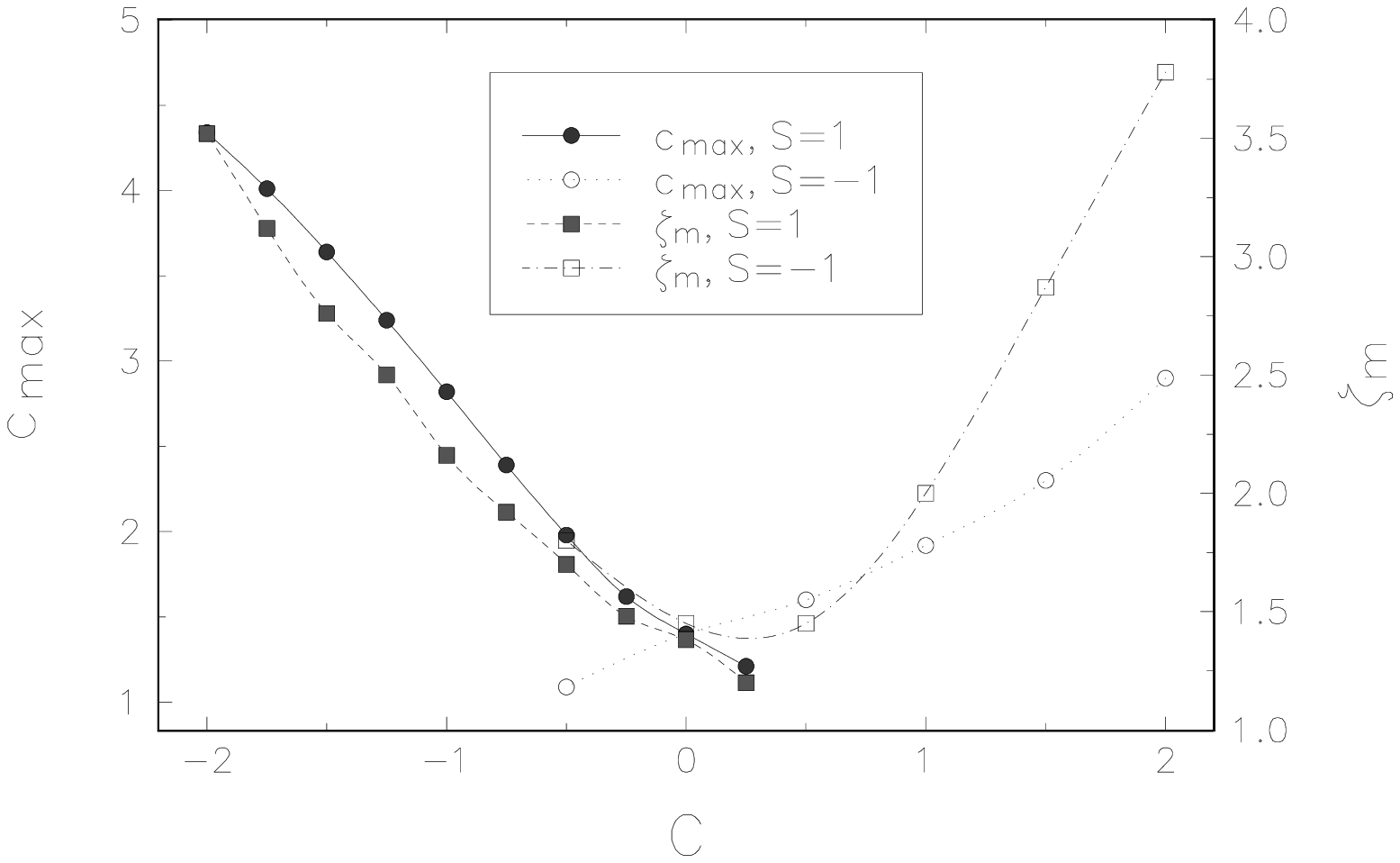}}
\caption{The maximal compression factor, $c_{max}$, and the 
distance to the point of the minimal pulse width,
 $\zeta_m$, as a function of the initial spatial and 
temporal pulse chirps, $C_{\xi}=C$ \, and \, $C_{\tau }=\pm C$, 
respectivelly. Spatial focusing chirp occurs for $C<0$,
temporal focusing chirp occurs for $C_{\tau}<0$, $\sigma=-0.5$ 
and $N^2=1.0$.}
\label{fcomp}
\end{figure}

\begin{figure}
\raisebox{-1cm}[7cm][2cm]{
\hspace*{-2cm}
\epsffile{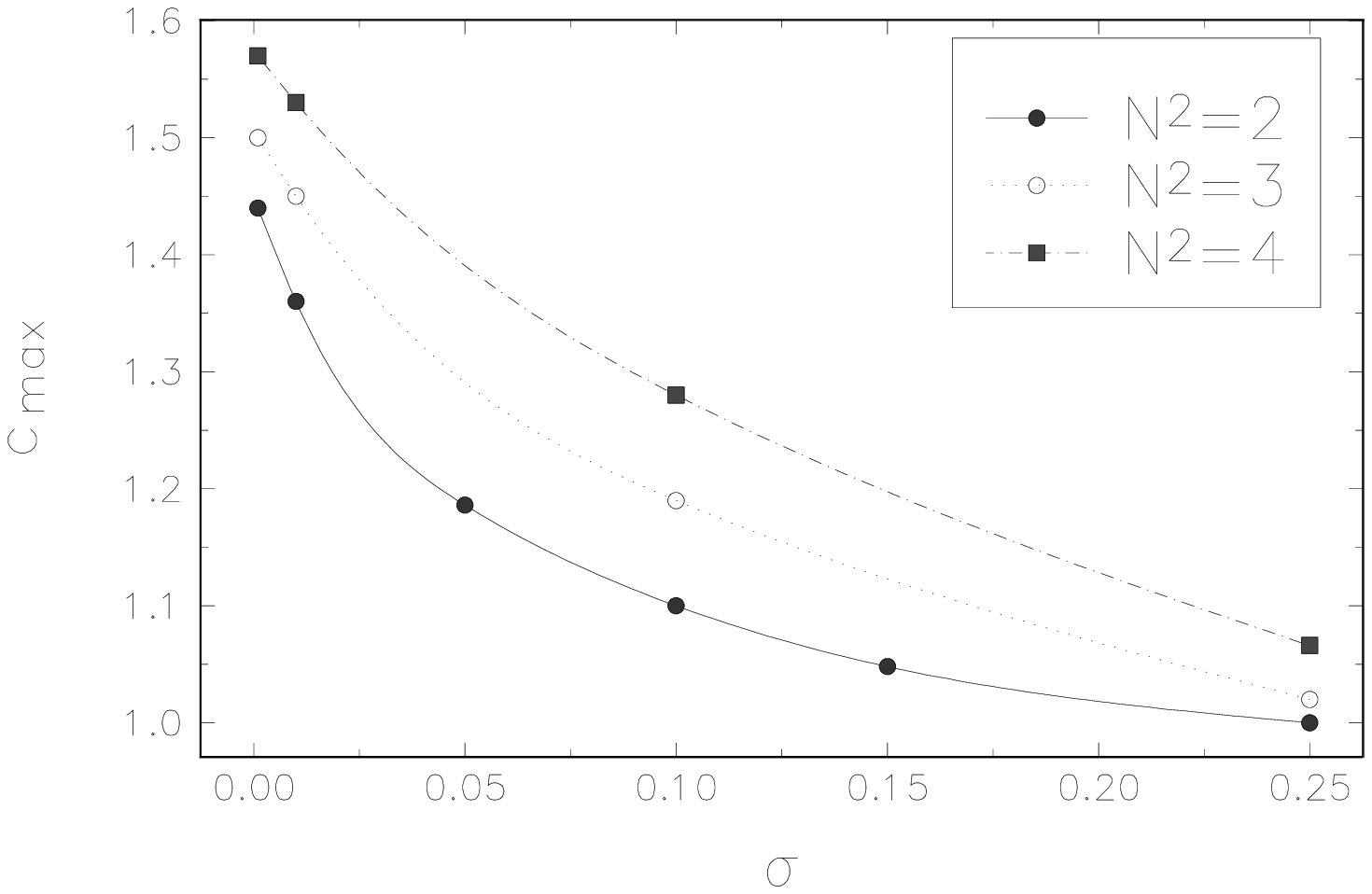}}
\caption{The maximal compression factor, $c_{max}$, as a 
function of the relative strength of dispersion and diffraction, 
$\sigma>0$, for different value of the strength of nonlinearity, 
$N^2$, and for initial Gaussian pulse with flat phase front.}
\label{ncomp}
\end{figure}

\begin{figure}
\raisebox{-1cm}[7cm][2cm]{
\hspace*{-2cm}
\epsffile{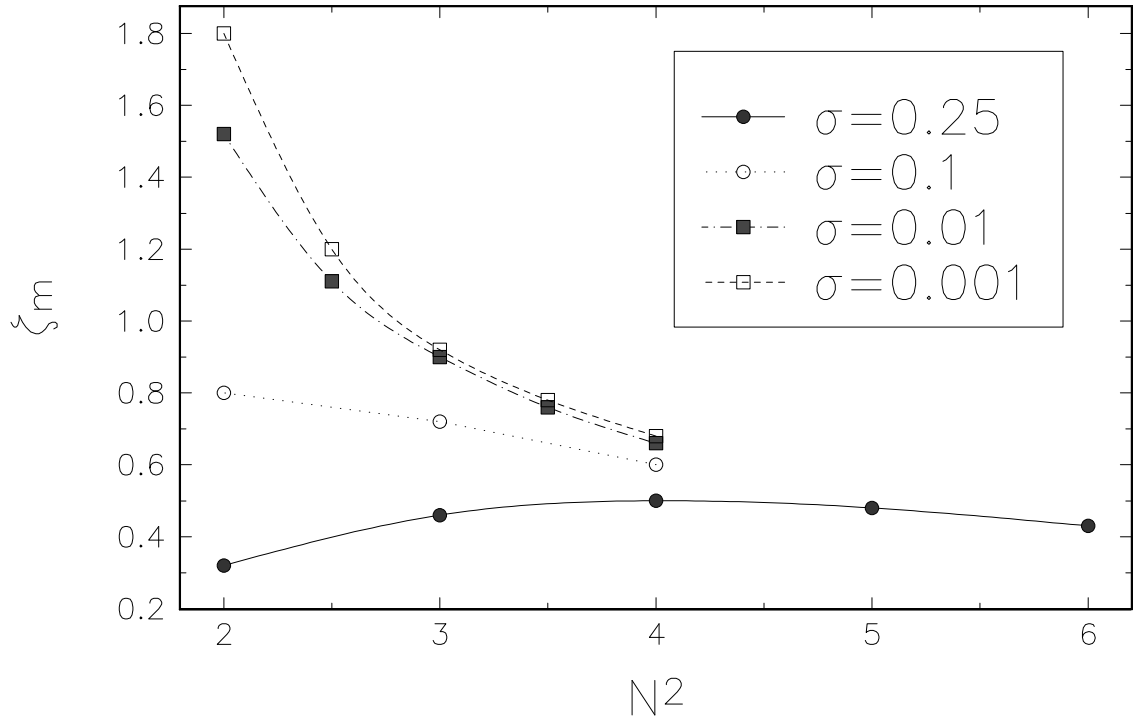}}
\caption{The distance to the point of the minimal pulse width, 
$\zeta_m$,  as a function of the strength of nonlinearity, $N^2$, 
for different value of the relative strength of dispersion and 
diffraction, $\sigma>0$, 
and for initial Gaussian pulse with flat phase front.}
\label{nodl}
\end{figure}

\begin{figure}
\raisebox{-1cm}[7cm][2cm]{
\hspace*{-2cm}
\epsffile{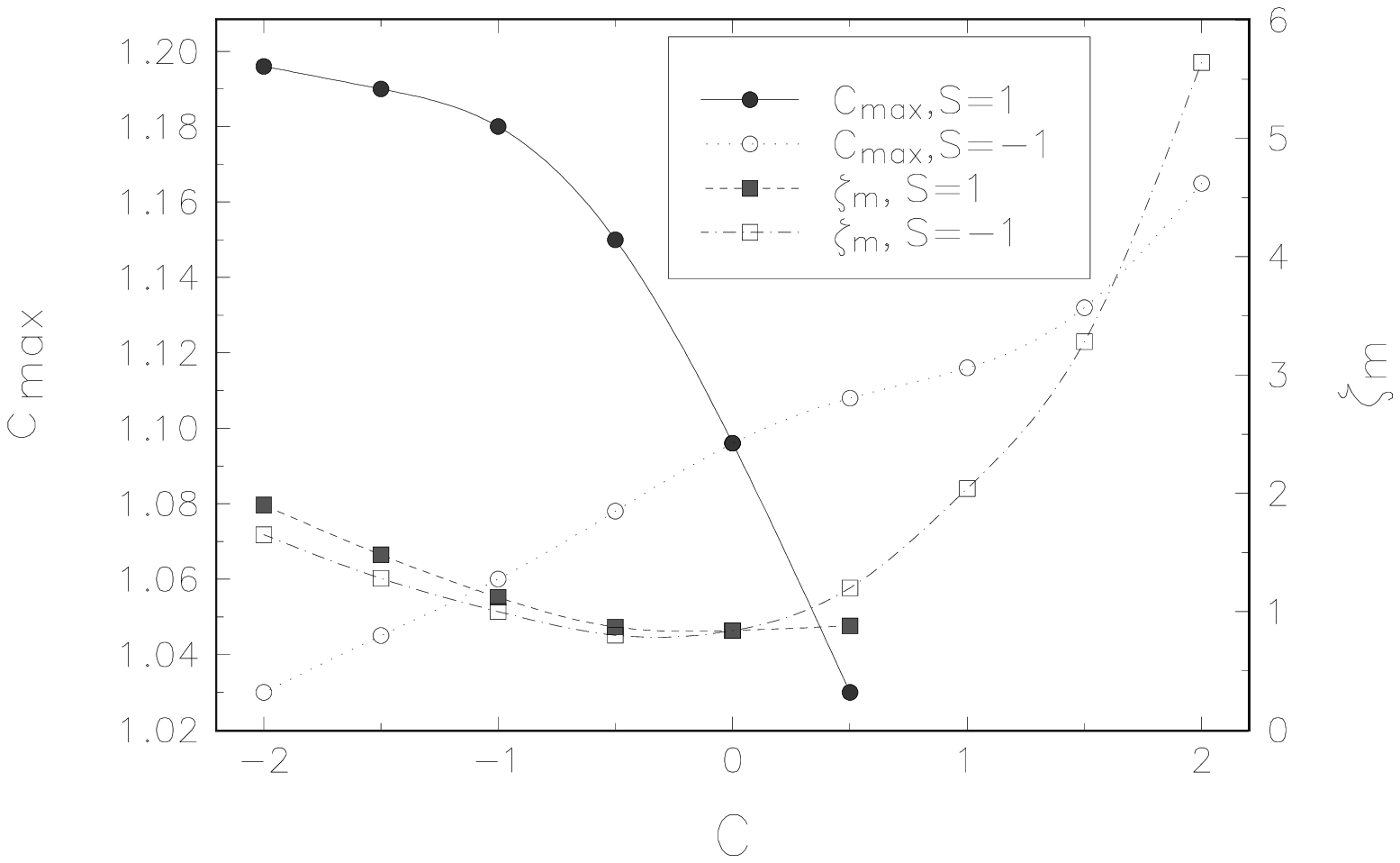}}
\caption{The maximal compression factor, $c_{max}$, and the 
distance to the point of the minimal pulse width, 
$\zeta_m$, as a function of the initial spatial and 
temporal pulse chirps, $C_{\xi}=C$ \, and \, $C_{\tau }=\pm C$,
respectivelly. Spatial focusing chirp occurs for $C_{\xi}<0$, 
temporal focusing chirp occurs for $C_{\tau}>0$, 
$\sigma=0.1$ and $N^2=2.0$.}
\label{fcomp}
\end{figure}

\end{document}